# Dynamics of warped galaxies

By ROBERT W. NELSON[1] AND SCOTT TREMAINE[1,2]

[1]Canadian Institute for Theoretical Astrophysics, McLennan Labs, University of Toronto, 60 St. George St., Toronto M5S 1A7, Canada

[2]Institute of Astronomy, Madingley Road, Cambridge CB3 0HA, UK

Large-scale warps in the outer parts of spiral galaxy discs have been observed for almost forty years, but their origin remains obscure. We review the dynamics of warped galaxy discs. We identify several mechanisms that could excite warps, all involving the gravitational interaction between the disc and the dark-matter halo.

## 1. Introduction

The HI gas in the outer regions of spiral galaxies commonly rises above and below the inner galactic plane in characteristic 'integral sign' warps. Such bending distortions were first observed nearly 40 years ago in our own galaxy (Burke 1957, Kerr 1957; see Burton 1988 for a review), and since then the nature and excitation of warped discs has been a major unsolved problem in galactic dynamics (Binney 1992). Interest in warped galaxies has increased with the recognition that they are quite common: all three spiral galaxies in the Local Group are warped, and most edge-on disc galaxies appear to be warped to some extent (Sánchez-Saavedra et al. 1990, Bosma 1991). The high frequency of warps implies that they are either long-lived or repeatedly excited.

Briggs (1990) summarized the characteristics of 12 isolated galaxies with HI warps. He found that the warp typically begins near the outer edge of the stellar disc, between $R_{25}$ (at the 25$^{\text{th}}$ mag sec$^{-2}$ isophote) and the Holmberg radius, $R_{26.5}$. The line of nodes is straight out to $R_{26.5}$, and then curves into a loosely wound leading spiral at larger radii.

Warps are most prominent in the HI gas disc, which extends to much larger radii than the stellar disc. In fact the HI warp typically begins just where the stellar disc ends, while inside this radius both the stellar and gas discs are remarkably flat. Evidence for warped stellar discs is seen in a number of nearby galaxies (e.g. M31 [Arp 1964, Innanen et al. 1982], NGC 4244 [van der Kruit 1979], M33 [Sandage & Humphreys 1980], NGC 4565 [van der Kruit & Searle 1981], NGC 5907 [Sasaki 1987], UGC 1770 [Goad & Roberts 1981], UGC 7170 [Cox et al. 1995], the Galaxy [Freudenreich et al. 1994]), but in some cases the apparent warp may simply reflect spiral structure in an imperfectly edge-on disc, and in most cases the stellar warp is not at all prominent.

Although this review concentrates on the gravitational dynamics of warps, there is little direct evidence that warps are a gravitational phenomenon that affects star and gas discs equally. It is possible that warps are generated by hydrodynamic or hydromagnetic forces that act only on the gas, although no compelling mechanism has been advanced (e.g. Binney 1991). The modest warps in the stellar discs would then represent the response of the disc stars to the gravitational perturbation from the warped gas. To test this possibility we may ask the questions: (i) do gas and old stars at the same radius suffer the same warp† (Sasaki 1987 has modelled the warp in NGC 5907 and concludes that the star and gas discs are warped by the same amount at the same radius)? (ii) do edge-on gas-free disc galaxies exhibit warps (the edge-on S0 galaxy NGC 4762 [Sandage 1961] shows a marginal warp in its stellar disc)?

---

† Warping in young stars (e.g. Porcel & Battaner 1995) is largely irrelevant since one might expect recent star formation in the warped gas.

Roughly half of warps are asymmetric (Bosma & Athanassoula 1990). Moreover, at least half of all galaxies, with or without warps, exhibit asymmetric outer HI distributions (Baldwin *et al.* 1980, Richter & Sancisi 1994). Although we discuss only warps in this review, the origin of these asymmetries is also puzzling and may have a related explanation.

## 2. Warps as normal modes

The concept that warps are free modes of oscillation of the galactic disc is due to Lynden-Bell (1965), and as in many other subjects we continue to explore the implications of his interesting idea. Lynden-Bell's suggestion was elaborated by Hunter & Toomre (1969), who investigated bending modes of isolated, cold, zero-thickness, self-gravitating discs. They found that (i) all axisymmetric or bowl-shaped ($m = 0$) and integral-sign ($m = 1$) warp modes are stable; (ii) discs with realistically smooth outer edges have no discrete bending modes at all (other than a trivial tilt), but instead support only a continuum of bending modes. Conclusion (ii) implies that any packet of bending waves should disperse in much less than a Hubble time, so the survival of warps is a puzzle. Hunter and Toomre's work preceded the recognition that galaxies have massive dark halos (J. Ostriker & Peebles 1973); thus it was only much later that Dekel & Shlosman (1983) and Toomre (1983a) recognized that the trivial neutral tilt mode of an isolated disc becomes nicely warped—and remains discrete—if the disc is embedded in the static potential of a flattened, rigid halo. Thus warps may arise from misalignment between the disc and the symmetry planes of the halo, a common feature in simulations of hierarchical galaxy formation (Katz & Gunn 1991). The dynamics of this 'modified tilt mode' have been analyzed in the linearized approximation by Sparke & Casertano (1988), who argue that this mode is consistent with the general shape of warps and the observation that the line of nodes of the warp is straight inside the Holmberg radius; even Briggs' (1990) observation that the line of nodes curls into a leading spiral at larger radii is a natural consequence of settling of the disc to a single mode as other waves propagate to the edge of the disc (Hofner & Sparke 1994). Kuijken (1991) discusses the nonlinear tilt mode.

### 2.1. *Dynamical friction*

One concern with this model is that real galaxy halos are not rigid. The collisionless halo material responds to the time-varying gravitational field of a warped, precessing disc, transferring energy and angular momentum between the disc and the halo and thereby damping (or possibly exciting) the normal mode (Toomre 1983a, Dekel & Shlosman 1983, Binney 1992). This process is simply dynamical friction in a novel context; unfortunately neither the timescale nor even the sign of the energy exchange from disc to halo is obvious.

The effects of dynamical friction on warps have now been addressed in two separate calculations. Nelson & Tremaine (1995) used perturbation theory (Lynden-Bell & Kalnajs 1972) to compute the rate of energy transfer to a spherical halo from a warped disc that precesses with pattern speed $\Omega_p$. They find that a non-rotating halo in which the distribution function is a decreasing function of energy alone always removes energy from the warp. A halo with predominantly radial orbits removes energy more rapidly than an isotropic halo, while in the unlikely case that the halo orbits are predominantly circular, the halo can add energy to the warp. The effect of halo rotation is generally to reduce the rate of energy loss if the halo rotates in the same direction that the warp precesses, and to increase the rate of energy loss otherwise; in some cases a rotating halo can even add energy to the warp.

The effects of halo rotation can be estimated by the following heuristic argument. Consider a halo rotating at uniform angular speed $\Omega_h$, surrounding a warp with pattern speed $\Omega_p$. The energy of the warp in the halo frame, $E_r$, is related to the energy in an inertial frame by $E_r = E - \Omega_h J_z$, where $J_z$ is the $z$-component of the angular momentum. Dynamical friction from the rotating halo generally removes energy from the warp in the rotating frame, in other words $\dot{E}_r < 0$ (this statement is exact if the halo distribution function is a decreasing function of $E_r$ alone). Moreover we expect that $\dot{E}_r = 0$ if the warp corotates with the halo ($\Omega_p = \Omega_h$); thus it is plausible that $\dot{E}_r = -k(\Omega_p - \Omega_h)^2 + O[(\Omega_p - \Omega_h)^3]$, where $k$ is a constant. The energy and angular momentum of the warp are related by $E = \Omega_p J_z$. Eliminating $J_z$, we have $\dot{E} = \dot{E}_r(1 - \Omega_h/\Omega_p)^{-1} = -k\Omega_p(\Omega_p - \Omega_h) + O[(\Omega_p - \Omega_h)^2]$. Neglecting the higher order terms, we find that energy is added to the warp by a rotating halo if and only if $\Omega_h/\Omega_p > 1$.

Whether energy transfer to the halo damps or excites a warp depends on the sign of the warp energy. We evaluate the warp energy in the next subsection; equation (2.7) shows that if the pattern speed is slow, $|\Omega_p| \ll \Omega$, then the energy of the warp is positive for retrograde warps ($\Omega_p < 0$) and negative for prograde warps ($\Omega_p > 0$). In the usual situation where the unperturbed disc is nearly aligned with the equatorial plane of an axisymmetric oblate corotating halo, the warp is retrograde (for the same reason that the line of nodes of a low-inclination near-Earth satellite regresses), so the energy is positive.

These results show that in the most likely situations (an oblate halo with isotropic or mainly radial dispersion tensor, which is non-rotating or rotating in the same direction as the disc) dynamical friction damps a warp. In unusual situations (nearly tangential dispersion tensor, prolate halo, or halo rotating in the opposite direction to the disc) friction can excite the warp (something like this possibility was first suggested by Bertin & Mark 1980; however their WKB analysis neglected the halo's contribution to the vertical restoring force and therefore did not recognize that unusual configurations were required for excitation). Of these possibilities for warp excitation, the most plausible is the counter-rotating disc and halo, a situation that could easily occur depending on the formation and merger history.

Nelson & Tremaine (1995) also find that the characteristic damping or excitation time $|E/\dot{E}|$ is generally much less than a Hubble time. A typical model has an oblate halo (axis ratio of the equipotential surfaces 0.8–0.9) with predominantly radial orbits ($\beta \equiv 1 - \langle v_\theta^2 \rangle/\langle v_r^2 \rangle = 0.3$, similar to values seen in simulations of halo formation) and circular speed $v_c = 200\,\mathrm{km\,s^{-1}}$; the disc is exponential with scale length $R_d = 4\,\mathrm{kpc}$ and total mass $6 \times 10^{10}\,\mathrm{M}_\odot$; and the warp shape resembles the warp in our own Galaxy. In this situation the damping time is less than $10^8\,\mathrm{yr}$—less than one orbital period of the disc. The damping time is longer if the halo dispersion tensor is isotropic ($\beta = 0$), or if the warp is only in the gaseous component of the disc, but it is difficult to find a plausible warp that survives unchanged for a Hubble time.

Dubinski & Kuijken (1995) have examined the evolution of tilted discs (both rigid and N-body) embedded in N-body halos. When the disc mass is comparable to the halo mass within the disc radius, the disc and halo align in a few orbital times—a result qualitatively similar to the analytic calculations. The simulations also show how the damping occurs: the inner halo twists so that its equatorial plane is aligned with the disc (the disc contains most of the angular momentum, so the halo follows the disc rather than vice versa). The common inclination of the disc and inner halo decays more slowly through friction from the outer halo; an 'alignment wave' propagates slowly out through the halo. Dubinski & Kuijken (1995) estimate that within one Hubble time dynamical friction will align the disc and the halo out to a radius of $\sim 100\,\mathrm{kpc}$.

The short damping/excitation times obtained in these complementary analytic and numerical investigations imply that warps cannot be primordial; they must be recently or continuously excited. Nevertheless, warps may be well approximated as a single discrete normal mode, since the combined excitation and damping mechanisms may excite the nearly neutral modified tilt mode to larger amplitude than the others.

### 2.2. *Negative-energy modes*

Negative-energy modes are a common feature of rotating systems (Pierce 1974). When dynamical friction removes energy and angular momentum from a negative-energy bending mode, the mode is excited rather than damped.

It is instructive to derive the energy of a bending mode. The equation of motion for small vertical displacements $Z(R,\phi,t)$ of a cold, axisymmetric, razor-thin disc is (Hunter & Toomre 1969)

$$\frac{D^2 Z}{Dt^2} = \left(\frac{\partial}{\partial t} + \Omega \frac{\partial}{\partial \phi}\right)^2 Z(\mathbf{R},t) = -\nu_h^2(R) Z(\mathbf{R},t) - G \int d\mathbf{R}' \mu(R') \frac{[Z(\mathbf{R},t) - Z(\mathbf{R}',t)]}{|\mathbf{R}-\mathbf{R}'|^3}. \tag{2.1}$$

Here $\mathbf{R} = (R,\phi)$ denotes position in the disc plane; $\Omega(R) > 0$ and $\mu(R)$ are the angular speed and surface density of the disc; and $-\nu_h^2(R)Z$ is the restoring force from the halo. This equation can be derived from the variation of an action, $\delta \int \mathcal{L}(Z,Z_t) d\mathbf{R} dt$, where $Z_t \equiv \partial Z/\partial t$. The Lagrangian density

$$\mathcal{L} = \tfrac{1}{2}\mu(R)\left[Z_t^2 + 2\Omega(R) Z_t Z_\phi + \Omega^2(R) Z_\phi^2 - \nu_h^2(R) Z^2\right] - \frac{G\mu(R)}{4} \int d\mathbf{R}' \mu(R') \frac{(Z-Z')^2}{|\mathbf{R}-\mathbf{R}'|^3}. \tag{2.2}$$

The canonical momentum density is then

$$p \equiv \frac{\partial \mathcal{L}}{\partial Z_t} = \mu(R)\left[Z_t + \Omega(R) Z_\phi\right], \tag{2.3}$$

and the energy $E_w = \int d\mathbf{R}(pZ_t - \mathcal{L})$ is given by

$$E_w = \tfrac{1}{2} \int d\mathbf{R}\mu(R)\left(Z_t^2 - \Omega^2 Z_\phi^2 + \nu_h^2 Z^2\right) + \tfrac{1}{4} G \int d\mathbf{R} d\mathbf{R}' \mu(R)\mu(R') \frac{(Z-Z')^2}{|\mathbf{R}-\mathbf{R}'|^3}. \tag{2.4}$$

Multiplying (2.1) by $Z$ and integrating over $\mu(R)d\mathbf{R}$ yields

$$0 = \int d\mathbf{R}\mu(R)\left(ZZ_{tt} + 2\Omega Z Z_{\phi t} + \Omega^2 Z Z_{\phi\phi} + \nu_h^2 Z^2\right) + \tfrac{1}{2} G \int d\mathbf{R} d\mathbf{R}' \mu(R)\mu(R') \frac{(Z-Z')^2}{|\mathbf{R}-\mathbf{R}'|^3}. \tag{2.5}$$

Subtracting $\tfrac{1}{2}\times$(2.5) from (2.4) we obtain an alternative expression for the energy,

$$E_w = \tfrac{1}{2} \int d\mathbf{R}\mu(R)\left[Z_t^2 - 2\Omega(R) Z Z_{\phi t} - Z Z_{tt}\right], \tag{2.6}$$

which is equation (28) of Nelson & Tremaine (1995). Although (2.6) is simpler than (2.4), the integrand in (2.4) has a simple physical meaning (the energy density) while the integrand in (2.6) does not.

For a mode in which $Z(R,\phi,t) = Z(R)\cos[m(\phi - \Omega_p t) + \psi(R)]$, the energy is

$$E_w = \pi m^2 \Omega_p \int R dR \mu(R) Z^2(R)[\Omega_p - \Omega(R)]. \tag{2.7}$$

Thus non-axisymmetric modes have positive energy if the pattern speed $\Omega_p$ is negative;

modes have negative energy if the pattern speed is positive but less than a suitable average of $\Omega(R)$, weighted by the surface density and mode amplitude.

Excitation of negative-energy modes by dynamical friction is an example of secular instability. The presence of secular instability depends on the nature of the dissipative forces. In this respect, dynamical friction closely resembles the gravitational radiation reaction force: both remove energy and angular momentum in the ratio $\Omega_p$, and both arise from the gradient of a potential and hence conserve circulation on constant-entropy surfaces in a fluid system. For either dissipative force, the criterion for secular instability in the present context is the existence of one or more negative-energy modes, $E_w < 0$ (Lynden-Bell & J. Ostriker 1967, Friedman & Schutz 1978).†

Modes with $m = 0$ or $m = 1$ have positive energy in isolated, cold, zero-thickness discs (Hunter & Toomre 1969); in the presence of a halo ($\nu_h^2 > 0$) the $m = 0$ modes still have positive energy and the $m = 1$ modes generally have positive energy if the halo is oblate (Nelson & Tremaine 1995). However, modes with $m > 1$ can and do have negative energy.

The ubiquity of negative-energy modes in rotating fluids is the basis for the famous result (Friedman & Schutz 1978) that *all* rotating stars are secularly unstable in the presence of gravitational radiation reaction force for sufficiently large azimuthal wavenumbers $m$ (although in most cases the growth rate of the instability is extremely small). Similarly, all cold, razor-thin, axisymmetric discs have negative-energy modes, which are secularly unstable in the presence of dynamical friction (e.g. figure 1).

Thus we face an uncomfortable situation: for most plausible halo models, positive-energy modes should be damped and negative-energy modes excited. Yet in the plausible case that the halo is oblate, the only bending modes that we detect ($m = 1$) have positive energy, while there is no sign of the many negative-energy modes that we know to be present.

Why are negative-energy modes not excited to observable amplitudes? One possibility is that halos add energy to warps, for example because the disc and halo always counter-rotate. However, there is a less dramatic explanation.

All bending waves other than the rare discrete modes (such as the modified tilt mode) eventually propagate to the edge of the disc, where they are damped by nonlinear or dissipative effects (Hunter & Toomre 1969). Thus negative-energy disturbances can only grow significantly if the growth time is less than the propagation time. These characteristic times are easy to estimate in the WKB approximation. The amplitude of a WKB bending wave subject to dynamical friction from the halo grows at a rate $\exp(\gamma t)$, where (Bertin & Mark 1980, eq. III-18 as $\zeta_0 \to 0$, with $k^2$ replaced by $k^2 + m^2/R^2$ to allow for large azimuthal wavenumbers)

$$\gamma = \frac{2^{5/2}\pi^{3/2}G^2\rho_h\mu\Omega_p}{3(k^2 + m^2/R^2)\sigma^3(\Omega - \Omega_p)}. \qquad (2.8)$$

Here $k$ is the radial wavenumber, $\rho_h$ and $\sigma$ are the density and one-dimensional velocity dispersion of the halo, and as usual $\mu$ and $\Omega$ are the surface density and angular speed

---

† The energy in (2.4) is Friedman & Schutz's canonical energy functional. The secular stability criterion of Lynden-Bell & J. Ostriker (1967) is equivalent to the condition that (2.4) is positive for all displacements $Z$ when $Z_t = 0$. The Lynden-Bell & Ostriker criterion can give incorrect results, as discussed by Hunter (1977) and Friedman & Schutz (1978), but turns out to be correct if the dissipative force is dynamical friction. Other dissipative forces, such as viscosity in fluid discs or shocks induced by density waves (Kalnajs 1972), lead to different stability criteria.

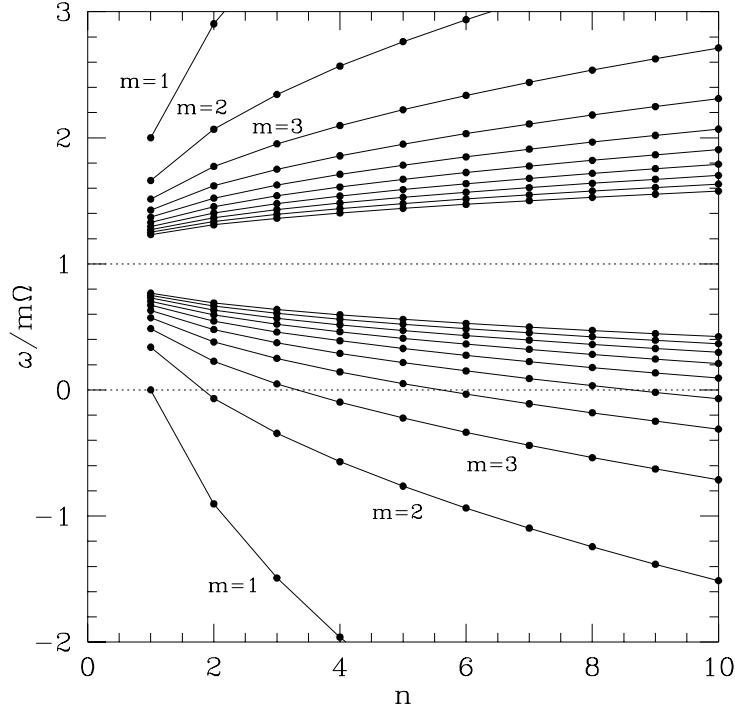

FIGURE 1. The frequencies of the bending modes of a cold, razor-thin disc that rotates at constant angular speed $\Omega$ (Bryan 1889, Hunter & Toomre 1969, Polyachenko 1977). The modes are labelled by the azimuthal wavenumber $m$ and the radial quantum number $n > 0$; modes with $n \leq 10$ and $0 < |m| \leq 10$ are shown. The vertical displacement in mode $(n, m)$ is $Z(R, \phi, t) = \xi^{-1} P_{2n+m-1}^m(\xi) \exp[i(m\phi - \omega t)]$, where $\xi^2 = 1 - R^2/a^2$ and $a$ is the outer radius of the disc. According to (2.7) modes have negative energy if and only if $0 < \omega/(m\Omega) < 1$, corresponding to the region between the dotted lines; there are many negative-energy modes with $m > 1$.

of the disc. The group velocity of the bending wave is

$$c_g = -\frac{\pi G \mu \, \text{sign}(k)}{m(\Omega - \Omega_p)}. \tag{2.9}$$

Thus the amplitude of a wavepacket grows by a factor $e$ over a radial distance

$$L \equiv \left|\frac{c_g}{\gamma}\right| = \frac{3(k^2 + m^2/R^2)\sigma^3}{2^{5/2}\pi^{1/2} G \rho_h m |\Omega_p|}. \tag{2.10}$$

If the halo is a singular isothermal sphere, then $\rho_h = \sigma^2/(2\pi G R^2)$, and we may write

$$\frac{L}{R} = \frac{3\pi^{1/2}}{2^{3/2}} [(kR)^2 + m^2] \left(\frac{\sigma}{\Omega R}\right) \left(\frac{\Omega}{m|\Omega_p|}\right). \tag{2.11}$$

Significant excitation requires $L/R \lesssim 1$; since $\sigma \simeq \Omega R$ and negative-energy waves have $0 < \Omega_p < \Omega$, this condition requires in turn $[(kR)^2 + m^2]/m \lesssim 1$, which requires $|kR| \lesssim \frac{1}{2}$ even for the most favorable value of $m$. This is difficult to achieve since $|kR| \gtrsim 2$ for even the fundamental mode in a typical disc. Thus excitation of negative-energy modes by dynamical friction from the halo is not likely to be important.

### 2.3. *Internal disc dynamics*

All of the results described so far are for cold and razor-thin warped discs. These do not accurately represent all of the important dynamics in real galaxy discs; for example, they are subject to violent axisymmetric, planar instabilities—which traditionally are ignored because planar motions are decoupled from the bending motions at linear order.

We may investigate waves in more realistic discs using a planar distribution of stars that is infinite and homogeneous in the $x$- and $y$-directions and held together by self-gravity in the $z$ or vertical direction, with a velocity-dispersion tensor that is independent of position. We consider the anisotropic distribution function (Spitzer 1942, Camm 1950)

$$F(z, \mathbf{v}) = \frac{\rho_0}{(2\pi)^{3/2} \sigma^2 \sigma_z} \exp\left(-\frac{v_x^2 + v_y^2}{2\sigma^2} - \frac{E_z}{\sigma_z^2}\right), \qquad (2.12)$$

where the vertical energy $E_z = \frac{1}{2} v_z^2 + \Phi(z)$, $\Phi(z)$ is the unperturbed potential, and $\sigma$ and $\sigma_z$ are constants equal to the velocity dispersions in any horizontal direction and in $z$. With this distribution function, Poisson's equation becomes

$$\frac{d^2 \Phi}{dz^2} = 4\pi G \rho(z) = 4\pi G \rho_0 \exp\left(-\frac{\Phi}{\sigma_z^2}\right), \qquad (2.13)$$

which may be solved to yield

$$\rho(z) = \rho_0 \operatorname{sech}^2(z/h), \quad \Phi(z) = 2\sigma_z^2 \ln \cosh(z/h), \quad h \equiv \frac{\sigma_z}{(2\pi G \rho_0)^{1/2}} = \frac{\sigma_z^2}{\pi G \mu}, \qquad (2.14)$$

where the total surface density is

$$\mu = \int_{-\infty}^{\infty} \rho(z) dz = 2\rho_0 h. \qquad (2.15)$$

We consider a small disturbance whose dependence on time and the horizontal coordinates is $\exp[i(k_x x + k_y y - \omega t)]$; without loss of generality we can set $k_y = 0$, ignore motions in the $y$-direction, and drop the subscript on $k_x$. To investigate wave propagation requires solving the linearized Boltzmann and Poisson equations to find the dispersion relation $\omega(k)$ (Toomre 1966, 1983b; Araki 1985; Weinberg 1991; Toomre 1995).

The dispersion relation is analytic for a razor-thin disc ($\rho_0 \to \infty$, $\sigma_z \to 0$, $\mu =$constant) with non-zero horizontal dispersion ($\sigma \neq 0$), which we call Toomre's (1966) sheet. The vertical displacement $Z(x,t)$ is related to the acceleration of a particle with $x$-velocity $v$ by

$$\frac{d^2}{dt^2} Z(x_0 + vt, t) = \left(\frac{\partial}{\partial t} + v\frac{\partial}{\partial x}\right)^2 Z(x,t)\bigg|_{x=x_0+vt} = \left(-\omega^2 + 2v\omega k - k^2 v^2\right) Z(x_0 + vt, t). \qquad (2.16)$$

The average acceleration of all the stars in the disc at $(x,t)$ is therefore

$$\left(-\omega^2 - k^2 \sigma^2\right) Z(x,t). \qquad (2.17)$$

The vertical force per unit mass exerted on the disc by itself is easily shown to be (e.g. Toomre 1966, or Binney & Tremaine 1987, Problem 5-4)

$$F_z(x,t) = -2\pi G \mu |k| Z(x,t), \qquad (2.18)$$

and equating the force per unit mass to the average acceleration gives the dispersion relation of Toomre's sheet,

$$\omega^2 = 2\pi G \mu |k| - k^2 \sigma^2. \qquad (2.19)$$

The disturbance is stable if and only if $|k| < k_J \equiv 2\pi G \mu / \sigma^2$; this instability is reminiscent of the firehose or Kelvin-Helmholtz instability. (Coincidentally, the condition for stable disturbances parallel to the disc plane is simply $|k| > k_J$; thus $k_J$ is the Jeans wavenumber.)

The instability of the Toomre sheet should be suppressed in sheets with sufficient thickness; the sheet should already be stable when one radian of the wave exceeds the full thickness $2h$, in other words when $k_J \times 2h = 4\sigma_z^2/\sigma^2 \gtrsim 1$, or $\sigma_z/\sigma \lesssim 0.5$. This crude estimate has been refined by numerical solutions of the linearized Boltzmann equation, which show that the minimum dispersion ratio $R \equiv \sigma_z/\sigma$ required for stability is 0.30 (Toomre 1966) or more accurately 0.293 (Araki 1985), safely below the dispersion ratios $R = 0.5$–$0.6$ in real galactic discs.

Non-zero thickness also affects the propagation of bending waves, in at least two ways. First, the restoring force from self-gravity is smaller than in a razor-thin disc. The magnitude of this effect may be estimated by imagining that the vertical structure of the disc is not altered by the bending wave; in other words that in the presence of a bending wave $Z(x,t)$ the unperturbed density $\rho(z)$ simply becomes $\rho[z - Z(x,t)]$. Then the net vertical force per unit mass exerted by the disc on itself (eq. 2.18) is modified to

$$F_z(x,t) = -2\pi G\mu |k| T(|k|) Z(x,t), \quad \text{where} \quad T(u) = \frac{\int dz\,dz'\, \rho(z)\rho(z') \exp(-u|z-z'|)}{\left[\int dz\, \rho(z)\right]^2}. \tag{2.20}$$

A second effect of non-zero thickness is that bending waves can be damped—or perhaps excited—by wave-particle interactions (Landau damping): stars whose natural vertical oscillation frequency $\nu(E_z)$ resonates with the apparent (Doppler-shifted) frequency $\omega - kv$ of the bending wave can absorb energy from the wave, converting wave energy into random motion in the disc (this effect was already mentioned, though not calculated, by Hunter & Toomre 1969, p. 771).

Significant damping occurs if the Doppler-shifted frequency is within one or two times the 'thermal width' of the resonance,

$$\left|\frac{\operatorname{Re}(\omega) - \nu_c}{k\sigma}\right| \lesssim 1 \text{ or } 2, \tag{2.21}$$

where $\nu_c \equiv \nu(E_z = 0)$ is the frequency for small vertical oscillations. For example, consider a bending wave with wavenumber $\frac{1}{2}k_J$ and frequency $\omega = \pi G\mu/\sigma$ given by the dispersion relation (2.19) for Toomre's sheet. The central frequency of the sech$^2$ sheet is $\nu_c = 2^{1/2}\pi G\mu/\sigma_z$. From (2.21) we expect strong damping whenever $R \equiv \sigma_z/\sigma \gtrsim 0.5$–$0.7$. The damping is stronger for shorter wavelengths and weaker for longer waves.

Weinberg (1991) has determined the bending eigenmodes of a (slightly truncated) sech$^2$ sheet embedded in a halo potential by solving the linearized Boltzmann equation, and has also determined damping rates for propagating waves. In a disc with dispersion ratio $R \simeq 1$, he finds strong damping $[\operatorname{Im}(\omega)/\operatorname{Re}(\omega) < -(2\pi)^{-1}$, so the amplitude damps by more than a factor of $e$ in one period] when $k \gtrsim 0.2 k_J$. Toomre (1983b, 1995) has determined damping rates for bending waves in the sech$^2$ sheet as a function of $R$ and $k$, and finds (for example) that in discs with $R = 0.6$ there is strong damping for $k \gtrsim 0.2 k_J$.

For greater realism we can add to the sech$^2$ sheet a Coriolis force $\kappa \hat{e}_z \times \mathbf{v}$, so that the stars oscillate horizontally in a manner reminiscent of epicycle oscillations in a disc galaxy. This refinement does not affect the velocity dependence of the unperturbed distribution function (2.12) or the dispersion relation of the Toomre sheet (2.19) but does modify the dispersion relation and damping rates for discs with non-zero thickness. The strong damping remains present at the most relevant dispersion ratios, $R \simeq 0.6$, but weakens markedly near $R \simeq 1$ (Toomre 1983b, 1995).

What are the implications for warps? In the solar neighbourhood, where $\nu_c \simeq 2\pi/(6 \times$

$10^7$ yr and $\sigma \simeq 30\,\mathrm{km\,s^{-1}}$, even for low-frequency waves (2.21) predicts strong damping when $\lambda \lesssim 2$–$4\,\mathrm{kpc}$; thus all bending waves with wavelength less than a few kpc are rapidly erased. The damping is even faster if the wave has a significant non-zero frequency or is at larger radii where $\nu_c$ is smaller. Weinberg (1991) estimates that only waves with $\lambda \gtrsim 10\,\mathrm{kpc}$ can survive for a Hubble time.

In summary, long-wavelength bending waves are strongly damped or excited by dynamical friction from the halo, while short-wavelength waves are strongly damped by internal wave-particle interactions. In a realistic galaxy disc it is unlikely that *any* bending wave can survive for a Hubble time (except for the case of a counter-rotating disc and halo, in which dynamical friction excites bending waves).

## 3. Warp generation by gravitational noise

We have shown that galaxy halos damp—and occasionally excite—warps in much less than the Hubble time. Halos can also excite warps through stochastic gravitational forces ('gravitational noise'), which can arise from several possible sources:

(i) Halo black holes: Lacey & J. Ostriker (1985) have suggested that the dark matter in galactic halos may be composed of black holes of mass $M \approx 10^{6.5}\,\mathrm{M_\odot}$, arguing that such objects could explain the age-velocity dispersion relation in the Galactic disc. Within the disc radius there are $N \approx 10^{4.5}$ such objects, so the typical irregularity in the halo mass would be $\Delta M \approx N^{1/2} M \approx 10^9\,\mathrm{M_\odot}$. However, the existence of such objects is difficult to reconcile with the survival of the discs of Local Group dwarf galaxies (Rix & Lake 1993), globular clusters (Moore 1993) and the nuclear disc of M31 (Tremaine 1995); also, they may collect at the centers of galaxies and coalesce into single black holes that are inconsistent with observational constraints (Hut & Rees 1992; but see Xu & J. Ostriker 1994 for an opposing view); finally, heating by halo black holes cannot easily explain the radial dependence of the velocity dispersion in the disc (Lacey 1991). Constraints on halo black holes are reviewed by Carr (1994).

(ii) Halo clusters: Some of the problems with halo black holes are evaded if the halo is composed of clusters of dark objects such as planets, brown dwarfs, neutron stars, or stellar-mass black holes (Carr & Lacey 1987, Wasserman & Salpeter 1994, Moore & Silk 1995). Such clusters could have radii of a few pc and masses as large as $10^5$–$10^6\,\mathrm{M_\odot}$ (more if they comprise only a fraction of the halo). Globular clusters are the only known halo clusters, with $M \approx 10^5\,\mathrm{M_\odot}$ and $N \approx 100$; they contribute a typical irregularity $\Delta M \approx N^{1/2} M \approx 10^6\,\mathrm{M_\odot}$, too small to be interesting.

(iii) Dark matter infall: In most current cosmological models, low-mass halos form first, and then merge to form larger halos (e.g. Lacey & Cole 1993); this ongoing process stirs up the halo and hence could affect the disc. Most merging halos are disrupted far outside the disc; nevertheless, tidal forces from distant mergers, streams of dark matter from disrupted halos on nearly radial orbits, and occasional mergers with dense low-mass halos all contribute to the gravitational noise in the vicinity of the disc. Moreover, as we discuss in Section 5, the response of the inner halo can magnify gravitational noise generated in the outer halo. The amplitude of this low-level noise is difficult to measure in existing simulations of halo formation, since the effective particle mass is $\gtrsim 10^8\,\mathrm{M_\odot}$ (Dubinski & Carlberg 1991, Warren *et al.* 1992), which already contributes far more to the noise level than putative halo black holes or clusters.

(iv) Baryonic infall: Galaxies are believed to form by the dissipative collapse of baryons to the centers of dark halos (White & Rees 1978). The size of the collapsed baryonic component is determined by its angular momentum and hence is much smaller than the parent halo; thus in a typical merger the lower density halos disrupt first, leaving the higher density baryonic galaxies orbiting in a common halo. The galaxy merger occurs much later, after the galaxy orbits decay by dynamical friction from the halo. Theoretical estimates of the galaxy merger rate are still uncertain (Lacey & Cole 1993, Navarro *et al.* 1994); a more direct approach is based on the number density of interacting (and hence merging) galaxy pairs and an estimate of their lifetime. Toomre (1977) finds in this way that roughly 10% of disc galaxies have undergone major mergers, while Keel & Wu (1995) estimate 30%†. These estimates refer to mergers of galaxies of similar size; for a mass ratio $f$ both analytic arguments (Lacey & Cole 1993) and numerical simulations (Navarro *et al.* 1995) suggest that the number of mergers $N(>f) \propto f^{-1/2}$.

As a first step towards understanding the effects of gravitational noise from the halo, let us estimate the energy associated with the Galactic warp. The warp begins at about the solar radius $R_0 = 8.5$ kpc and its amplitude is given roughly by $h(R) = \max[0, \alpha(R-R_0)]$, with $\alpha \simeq 0.1$. Since the warp is seen mainly in HI we replace $\mu(R)$ by the surface density in HI, which is approximately $\mu_0 \exp[-(R - R_0)/R_d]$ where $\mu_0 = 10\,\mathrm{M_\odot\,pc^{-2}}$, and $R_d \simeq 0.5 R_0$ (Burton 1976, 1988). Inserting these parameters in (2.7), assuming a flat rotation curve with rotation speed $V_c$ and $|\Omega_p| \ll \Omega$, we find

$$E = -1 \times 10^7\,\mathrm{M_\odot} V_c^2 \left(\frac{\Omega_p R_d}{V_c}\right). \tag{3.22}$$

The pattern speed is poorly known but for a moderately oblate halo (axis ratio of isopotential surfaces 0.8–0.9) we expect $\Omega_p R_d/V_c \simeq -0.1$ (Nelson & Tremaine 1995). Thus $E \approx 1 \times 10^6\,\mathrm{M_\odot} V_c^2$; in other words *the energy in the Galactic warp is no larger than the kinetic energy of a single massive globular cluster.* The energy is small because (i) the HI mass outside the solar circle is small; (ii) the warp angle $\alpha$ is small and the energy is proportional to $\alpha^2$; (iii) the halo is not strongly flattened, so the pattern speed is small.

If the halo is composed of black holes of mass $M$, discrete positive-energy normal modes should be in equipartition‡ with the black holes; for an isothermal halo this implies $\langle E \rangle = kT = M\sigma^2 = \frac{1}{2}MV_c^2$, where $\sigma$ is the one-dimensional velocity dispersion and $V_c = 2^{1/2}\sigma$ for an isothermal sphere. Thus, if the warp is in equipartition, the observed amplitude implies $M \simeq 2 \times 10^6\,\mathrm{M_\odot}$, remarkably close to the value proposed by Lacey & J. Ostriker (1985) for other reasons.

Our estimate of the expected warp amplitude can be generalized approximately to the case where a fraction $f$ of the total halo mass within the disc radius is in $N$ objects of mass $M$. Denote the total halo mass by $M_h = NM/f$. The warp amplitude $h$ decays by dynamical friction at a rate proportional to the halo mass; thus $(dh/dt)_{\mathrm{fr}} = -CM_h h$, while gravitational noise from the massive objects excites the warp energy at a rate proportional to the number of objects and the square of their mass; thus $\frac{1}{2}(dh^2/dt)_n = DNM^2$. The equilibrium warp amplitude is therefore given by $h_{\mathrm{eq}}^2 = DNM^2/(CM_h) = DfM/C$. The warp energy $E \propto h^2$ so the equilibrium energy is $\propto fM$; when $f = 1$ the equilibrium energy is $\frac{1}{2}MV_c^2$ so in general

$$\langle E \rangle = \tfrac{1}{2}fMV_c^2 \simeq \frac{NM^2}{M_h}V_c^2 = \frac{\Delta M^2}{M_h}V_c^2, \tag{3.23}$$

---

† Tóth & J. Ostriker (1992) argue that most spiral galaxies cannot accrete more than a few per cent of their disc mass without excessive thickening of the disc. This limit may be too stringent, since it neglects such processes as tidal disruption of the infalling satellite, tilting (rather than heating) of the disc by the satellite (Huang & Carlberg 1995), and excitation of bending waves in the disc which are subsequently damped by the halo.

‡ In the absence of other damping mechanisms, discrete negative-energy modes would not come into equipartition, but would grow without limit, since they are excited both by stochastic gravitational noise and by dynamical friction.

where $\Delta M = N^{1/2} M$ is the typical irregularity in the halo mass averaged over a scale comparable to the disc size. Since $M_h \approx 10^{11}\,{\rm M}_\odot$ and $\langle E \rangle \approx 10^6\,{\rm M}_\odot V_c^2$, the noise level required to excite the observed warp is roughly $\Delta M \approx 10^{8.5}\,{\rm M}_\odot$. The required noise could be generated by halo black holes, dark clusters, dark matter infall, or baryonic infall, but not by globular clusters.

## 4. Warp generation by a twisting halo

Simulations of halo formation (e.g. Warren *et al.* 1992) show that most halos are triaxial and that the orientation of the principal axes remains roughly constant out to at least 50 kpc from the halo centre. The high degree of alignment probably reflects efficient radial transport by material on elongated orbits. However the orientation of this inner halo is not fixed: both analytic arguments and N-body simulations show that the orientation changes with each doubling of the halo age, in response to the infall of new halo material (see Binney 1991, 1992 for reviews).

E. Ostriker & Binney (1989) have suggested that warps represent the response of the disc to the changing orientation of the inner halo. Let us imagine that the orientation of the principal-axis system of the inner halo rotates relative to an inertial frame with angular speed $\Omega$ and that the inner disc lies in one of the principal planes (most likely the one perpendicular to the minor axis $\hat{\bf e}_z$). The frame rotation gives rise to a Coriolis force $-2\Omega \times {\bf v}$. The vertical component of this force may be written approximately as $-2\hat{\bf e}_z V_c(R) \Omega \cdot \hat{\bf e}_R$, where $V_c(R)$ is the circular speed at radius $R$ and $\hat{\bf e}_R$ is the unit polar vector in the disc plane. This vertical force distorts the outer disc into a warp; the amplitude of the warp is determined by the restoring force from the disc and the halo, and is greatest in the outer disc where the restoring forces are weakest. If we neglect the restoring force from the disc, the height of the warp should be given by

$$\frac{z(t)}{R} \simeq 0.07 \left[\frac{\Omega \cdot \hat{\bf e}_R}{10^{-10}\,{\rm yr}^{-1}}\right] \left(\frac{R}{15\,{\rm kpc}}\right) \left(\frac{220\,{\rm km\,s}^{-1}}{V_c}\right) \left(\frac{0.1}{\epsilon_\Phi}\right), \qquad (4.24)$$

where $\epsilon_\Phi$ is the ellipticity of the halo potential, and $V_c$ is taken from Kerr and Lynden-Bell (1986). This approximation is only valid at large radii, where the gravitational attraction from the massive inner disc is small; it mainly shows that a plausible rate of rotation of the halo principal axes (several radians per Hubble time) can produce a significant warp.

A shortcoming of this simple picture is that the halo is treated as rigid, whereas in fact Dubinski & Kuijken (1995) show that the inner halo (within 1–2 disc radii) tends to follow the disc rather than vice versa, because the disc has comparable mass but much more angular momentum, and hence is harder to re-orient.

The Ostriker-Binney proposal is distinct from mechanisms of warp generation based on normal modes because the warp is a forced oscillation rather than a free oscillation; thus damping by dynamical friction plays a much less important role. It is distinct from other mechanisms that rely on infall because the warp is caused by systematic rather than stochastic variations in the potential of the inner halo.

## 5. Warp generation by satellites

The tidal field from the Magellanic Clouds was investigated—and then discarded—as the cause of the warped disc of the Galaxy as soon as the warp was discovered (Burke 1957, Kerr 1957). The arguments are: (i) for any plausible mass the present tidal field is too weak by a factor of ten or more to excite the observed warp amplitude of several kpc (e.g. Hunter & Toomre 1969), and estimates of the Cloud orbit from its proper motion (Jones *et al.* 1994, Lin *et al.* 1995) and the dynamics of the Magellanic Stream (Murai & Fujimoto 1980, Lin & Lynden-Bell 1982) imply that its present distance is close to the minimum distance; (ii) warps appear in apparently isolated galaxies (Sancisi 1976) so either some satellites are invisible or satellites account for only some warps.

Argument (i) has been questioned by Weinberg (1995), who points out that estimates of the gravitational influence of the Clouds usually neglect the response of the Galactic halo, which can magnify the tidal field. Weinberg has computed the self-consistent response of the disc and halo to the tidal field from the Clouds. He finds that the disc warp is far larger than the direct tidal field would produce if no halo were present; given the uncertainties, an active halo distorted by the Clouds *may* be sufficient to excite the observed warp.

The following simple argument (Lynden-Bell 1985) suggests a reason for this strong enhancement. Consider a spherically symmetric halo with density $\rho(r)$ and enclosed mass $M(r) \equiv 4\pi \int_0^r \rho(x) x^2 \, dx$. The halo is subjected to a small potential perturbation $U^i(\mathbf{r}) = U^i_{\ell m}(r) Y_{\ell m}(\theta, \phi)$ ('i' for 'input'); this is the total perturbation, which includes both the external potential and the gravitational response of the halo. By solving the linearized Boltzmann equation we can determine the density response induced in the halo by $U^i$ and by solving the Poisson equation we can determine the potential perturbation caused by the density response, $U^o(\mathbf{r}) = U^o_{\ell m}(r) Y_{\ell m}(\theta, \phi)$ ('o' for 'output'). The input and output are related by a linear response operator,

$$U^o_{\ell m}(r) = \int_0^\infty dr' R_\ell(r, r') U^i_{\ell m}(r'). \qquad (5.25)$$

The input potential $U^i$ is the sum of the satellite potential $U^s$ and the halo response $U^o$.

We now construct a crude model for $R(r, r')$. We assume that the perturbation is static, which is incorrect for a satellite orbiting within the halo, but should approximately capture the response of the halo at radii much less than the satellite. The unperturbed potential at $r'$ is roughly $-GM(r')/r'$ so the fractional potential perturbation is $\xi(r') = -r' U^i(r')/GM(r')$. This should produce a fractional density perturbation of similar magnitude, so the perturbed mass in the interval $[r', r' + dr']$ will be $\delta M(r') = 4\pi r'^2 \lambda(r') \xi(r') \rho(r') dr'$, where $\lambda(r')$ is a dimensionless positive number that parametrizes the response of the halo (in practice, of course, stars have eccentric orbits and sample a range of radii, so $\delta M(r')$ depends on $U^i$ at a range of radii, but we neglect this complication). The response potential to $\delta M(r')$ is $U^o(r) = -4\pi G (2\ell + 1)^{-1} r^\ell_< r_>^{-\ell-1} \delta M(r')$, where $r_< = \min(r, r')$, $r_> = \max(r, r')$; since we are mostly interested in the response at small radii we use the simpler expression $U^o(r) = -4\pi G (2\ell + 1)^{-1} r^\ell r'^{-\ell-1} \delta M(r')$ for $r < r'$ and zero otherwise. Finally, we assume $\lambda$ is independent of radius since the dimensionless response of the halo should not depend strongly on position. Thus

$$R_\ell(r, r') = \frac{16\pi^2 \lambda_\ell}{2\ell + 1} \frac{r^\ell r'^{2-\ell} \rho(r')}{M(r')}, \qquad r < r', \qquad (5.26)$$

and zero otherwise.

Combining (5.25) and (5.26) we obtain

$$U^o_{\ell m}(r) = \frac{16\pi^2 \lambda_\ell r^\ell}{2\ell + 1} \int_r^\infty dr' \frac{r'^{2-\ell} \rho(r')}{M(r')} \left[ U^s_{\ell m}(r') + U^o_{\ell m}(r') \right], \qquad (5.27)$$

which is easily solved to yield

$$U^o_{\ell m}(r) = k_\ell r^\ell M(r)^{-k_\ell} \int_r^\infty \frac{U^s_{\ell m}(r')}{r'^\ell} M(r')^{k_\ell - 1} dM(r'), \qquad (5.28)$$

where $k_\ell = 4\pi\lambda_\ell/(2\ell+1)$. If the satellite is outside the halo, $U^s_{\ell m}(r) \propto r^\ell$ and the solution simplifies to

$$U^o_{\ell m}(r) = U^s_{\ell m}(r) \left[ \frac{M^{k_\ell}_{\rm tot}}{M(r)^{k_\ell}} - 1 \right], \tag{5.29}$$

where $M_{\rm tot}$ is the total halo mass. Thus the total potential $U^o + U^s$ is enhanced over the direct tidal potential $U^s$ by a factor $[M_{\rm tot}/M(r)]^{k_\ell}$.

Lynden-Bell (1985) derives $k_2 = 0.44$ if the halo is a singular isothermal gas sphere, but in general evaluating $k_\ell$ for a stellar system requires solving the linearized Boltzmann equation numerically. For the artificial case in which the stars are confined to travel on spheres, having a Maxwellian distribution of tangential velocities with rms tangential velocity equal to the local circular speed (Barnes *et al.* 1986), we find $k_1 = 0.9369$, $k_2 = 0.6162$, $k_3 = 0.4607$. In general we expect $0 < k_\ell \lesssim 1$. Thus, for example, if the halo mass inside the present distance of the Clouds is five times larger than the mass at 10 kpc, and $k_\ell \simeq 1$, the potential is enhanced by a factor of 5.

Calculations of the response of spherical stellar systems to static or slowly varying tidal potentials would be of considerable interest. In another context, this mechanism might also help large-scale features such as bars to concentrate gas at the centres of galaxies.

## 6. Conclusions

Galactic warps are so common that they must have a commonplace explanation.

Short-wavelength bending waves are damped by wave-particle resonances, while long $m = 1$ bending waves are strongly damped by dynamical friction from an oblate halo that is either non-rotating or rotating in the same direction as the disc. Thus warps cannot generally be primordial: their observed amplitudes must be determined by some ongoing or recent process.

A coherent $m = 1$ warp is the dominant response to a wide variety of excitations, because the modified tilt mode is nearly neutral and therefore can easily be driven to a large amplitude.

We have identified at least four mechanisms that can excite warps: (i) excitation by dynamical friction from the halo, if the halo and disc rotate in opposite directions; (ii) gravitational noise from the halo; (iii) Coriolis force from a twisting halo (the Ostriker-Binney mechanism); (iv) tidal fields from satellites, which are likely to be strongly magnified by the halo response. The distinction between these mechanisms is often blurry (e.g. the accretion of small satellites generates gravitational noise, and twists the principal axes of the halo), and less important than their common feature: all four are aspects of the time-varying gravitational field from an active and evolving halo.

Thus, although many details remain obscure, there is growing evidence that warps reflect the response of the outer disc to the complex dynamics of the galactic halo.

We thank Jerry Sellwood for many discussions, and Alar Toomre for access to his extensive unpublished notes and for thoughtful comments on an early draft of this review. ST thanks Donald Lynden-Bell for many insights and much wise advice over the past two decades, and for his hospitality during a research leave at Cambridge where much of this review was prepared.

This research was supported by a Killam Research Fellowship, a grant from the Raymond and Beverly Sackler Foundation, and by NSERC.